\documentclass[superscriptaddress,twocolumn,10pt,nofootinbib,floatfix]{revtex4-2}

\usepackage{amsfonts}
\usepackage{amssymb}
\usepackage{amsmath}
\usepackage{appendix}
\usepackage{times}
\usepackage{graphicx}
\usepackage{subeqnarray}
\usepackage{bbold}
\usepackage{enumerate}
\usepackage{color}
\usepackage{bm}
\usepackage[colorlinks=true, letterpaper=true, pdfstartview=FitV, linkcolor=blue, citecolor=blue, urlcolor=blue]{hyperref}

\usepackage[normalem]{ulem}

\setcitestyle{super,open={},close={}}

\makeatletter
\renewcommand\@biblabel[1]{#1.}
\makeatother

\begin{document}
\title{Chiral Altermagnetic Magnetoelectrics}

\author{Chengwu Xie}\thanks{C.X. and W.M. contributed equally to this work.}
\address{Institute for Superconducting and Electronic Materials, Faculty of Engineering and Information Sciences, University of Wollongong, Wollongong 2500, Australia.}
\address{Institute of Quantum Materials and Devices, School of Electronic and Information Engineering, \\ Tiangong University, Tianjin 300387, China.}

\author{Weizhen Meng}\thanks{C.X. and W.M. contributed equally to this work.}
\address{College of Physics, Hebei Key Laboratory of Photophysics Research and Application, Hebei Normal University, Shijiazhuang 050024, China.}

\author{Zhenzhou Guo}
\address{Institute for Superconducting and Electronic Materials, Faculty of Engineering and Information Sciences, University of Wollongong, Wollongong 2500, Australia.}

\author{Xiaodong Zhou}\email{zhouxiaodong@tiangong.edu.cn}
\address{Institute of Quantum Materials and Devices, School of Electronic and Information Engineering, \\ Tiangong University, Tianjin 300387, China.}
\address{School of Physical Science and Technology, Tiangong University, Tianjin 300387, China.}

\author{Shifeng Qian}
\email{qiansf@ahnu.edu.cn}
\affiliation{Anhui Province Key Laboratory for Control and Applications of Optoelectronic Information Materials, Department of Physics, Anhui Normal University, Wuhu 241000, Anhui, China.}

\author{Tie Yang}\email{yangtie@swu.edu.cn}
\address{School of Physical Science and Technology, Southwest University, Chongqing 400715, China}

\author{Wenhong Wang}
\address{Institute of Quantum Materials and Devices, School of Electronic and Information Engineering, \\ Tiangong University, Tianjin 300387, China.}

\author{Zhenxiang Cheng}
\address{Institute for Superconducting and Electronic Materials, Faculty of Engineering and Information Sciences, University of Wollongong, Wollongong 2500, Australia.}

\author{Xiaotian Wang}\email{xiaotianw@uow.edu.au}
\address{Institute for Superconducting and Electronic Materials, Faculty of Engineering and Information Sciences, University of Wollongong, Wollongong 2500, Australia.}

\begin{abstract}
In this work, we introduce a new class of chiral altermagnetic magnetoelectrics in structurally chiral, nonpolar altermagnetic systems and identify the experimentally well-characterized three-dimensional metal-organic framework K[Co(HCOO)$_3$] as a promising material platform. K[Co(HCOO)$_3$] exhibits chirality-locked \emph{g}-wave altermagnetic spin splitting together with dual-mode switchable electric polarization controlled by N\'eel-vector reorientation and structural chirality. Specifically, N\'eel-vector reorientation generates a finite electric polarization and reverses its sign, whereas chirality switching between left- and right-handed enantiomers produces an additional sign reversal. The associated electronic and optical responses provide effective readout channels for these switchable states. Our results establish chiral altermagnetic magnetoelectrics as a promising route to chirality- and N\'eel-vector-controlled nonvolatile multifunctional spintronics.
\end{abstract}

\maketitle

\textit{\textcolor{blue}{Introduction---}}Altermagnetism represents a third class of collinear magnetic order that uniquely bridges the fundamental dichotomy between conventional ferromagnetism and antiferromagnetism~\cite{add1,add2,guo_AM2025,add30,add31}. It is characterized by compensated magnetic moments and nonrelativistic alternating spin-split band structures~\cite{HY-Ma2021,RW-Zhang2024,add4,add5,add7,add8,FY-Zhang2025,B-Jiang2025}. This distinct characteristic enables altermagnets (AMs) to integrate the key advantages of ferromagnets (FMs) and antiferromagnets (AFMs), giving rise to a variety of emergent phenomena, including unconventional spin transport~\cite{add27,H-Bai2023,H-Bai2022,Karube2022}, anomalous Hall effect~\cite{add13,add14,add78,add80}, anomalous thermal transport~\cite{XD-Zhou2024,Badura2025,L-Han2024}, and magneto-optical effects~\cite{add58,Mazin2021,add19,add20,add21,add82,add84}. These features establish AMs as a promising platform for spintronic functionalities beyond the limits of conventional FMs and AFMs.

Among the various strategies for enriching altermagnetic functionalities, structural chirality provides a particularly attractive degree of freedom. Owing to the non-superimposability of a structure and its mirror image, the chiral crystal exists only in the 65 Sohncke space groups, which lack inversion, mirror, and roto-inversion symmetries~\cite{add32}. This distinctive structural property gives rise to a wide range of intriguing phenomena, including unconventional magnetotransport behavior~\cite{add33,add34,add35,add36}, quantum responses to circularly polarized light~\cite{add37,add38,add39}, and enhanced catalytic performance~\cite{add40,add41,add42}. Interestingly, the symmetry breaking inherent to chiral crystals also provides a favorable setting for unconventional couplings between electric and magnetic degrees of freedom~\cite{HN-Zheng2024}. Although substantial progress has been made in nonmagnetic, ferromagnetic, and antiferromagnetic chiral systems, chiral AMs remain largely unexplored.

Magnetoelectric materials, in which magnetic order induces electric polarization or electric polarization in turn influences magnetic order, provide a fundamental platform for coupling electric and magnetic degrees of freedom in solids. Recent studies on AMs have largely focused on systems where ferroelectric polarization is employed to manipulate spin-related properties~\cite{add24,add25,add26,YQ-Zhu2025,MQ-Gu2025,ZY-Zhu2025}. For instance, it has been shown that in certain hybrid improper ferroelectrics $\mathrm{[C(NH_2)_3]Cr(HCOO)_3}$~\cite{MQ-Gu2025}, the altermagnetic spin splitting is strongly coupled to the ferroelectric polarization, offering opportunities for electric-field control of spin states. Nevertheless, nonpolar AMs may enable electric polarization driven directly by N\'eel-vector reorientation, thereby providing a new pathway toward altermagnetic magnetoelectrics~\cite{XH-Tu2026,YX-Zhang2025,WT-Guo2025}.

Building on this idea, the coexistence of structural chirality and N\'eel-vector-driven polarization in nonpolar AMs provides a route to chiral altermagnetic magnetoelectrics, with dual-mode switchable electric polarization controlled by structural chirality and N\'eel-vector orientation. In this work, we construct a general theoretical model to establish this concept and further identify the synthesized three-dimensional (3D) metal-organic framework (MOF) K[Co(HCOO)$_3$]~\cite{add50} as a representative material platform. Specifically, the 3D MOF K[Co(HCOO)$_3$] with a hexagonal chiral lattice exhibits chirality-locked altermagnetic semiconducting properties in left/right-handed (C$_L$/C$_R$) enantiomers, accompanied by $\emph{g}$-wave spin splitting in 3D momentum space. Moreover, reorientation of the out-of-plane N{\'e}el vector generates a finite in-plane polarization $P_y$ in this nonpolar AM, following a $\sin(2\theta)$ dependence on the rotation angle $\theta$. Remarkably, this polarization can be switched in two distinct ways: N{\'e}el-vector reorientation within a fixed enantiomer reverses its sign, while switching between opposite enantiomers produces an additional sign reversal. The associated anomalous Hall and magneto-optical responses provide effective readout channels for these switchable states. These results establish K[Co(HCOO)$_3$] as a promising platform for chiral altermagnetic magnetoelectrics with coupled chirality, altermagnetism, and magnetoelectricity.

\begin{figure}[t]
\centering
\includegraphics[width=1\linewidth]{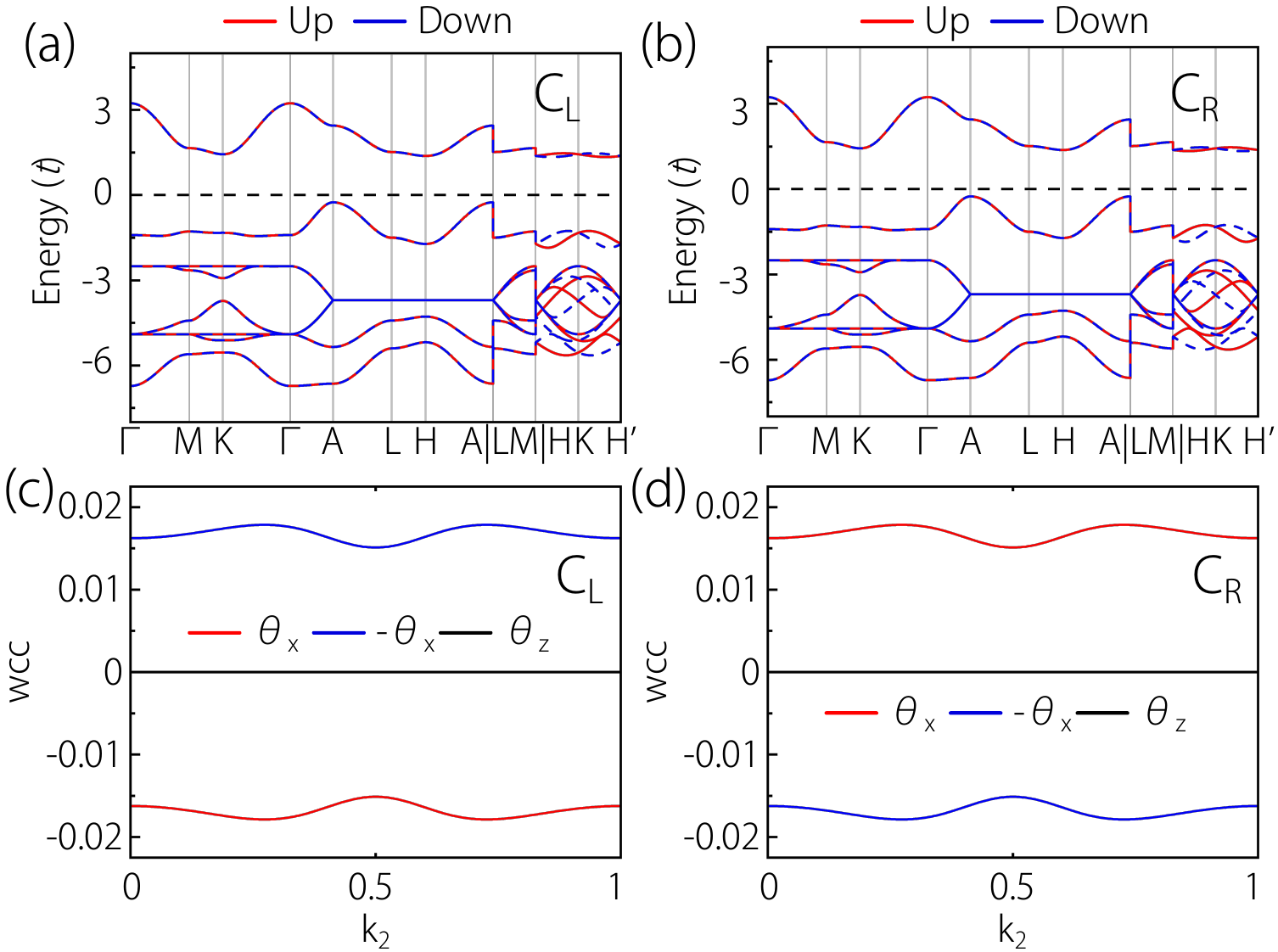}
\caption{Band structures of the (a) C$_L$ model and (b) C$_R$ model. Wilson loop of the occupied bands at $k_z=0$ for the (c) C$_L$ model and (d) C$_R$ model, with the N\'eel vector oriented along the $z$, $x$, and $-x$ axis. 
\label{fig1}}
\end{figure}

\textit{\textcolor{blue}{Symmetry principle for chiral altermagnetic magnetoelectrics---}}To uncover the fundamental physics underlying the interplay among structural chirality, altermagnetism, and magnetoelectricity, one may first establish the symmetry conditions that permit these distinct degrees of freedom to coexist. Because macroscopic responses and their cross-couplings are fundamentally governed by the crystalline and magnetic symmetries, we begin by formulating a general symmetry framework before turning to specific electronic structures or transport properties. This framework provides a general criterion for identifying and understanding chiral altermagnetic magnetoelectrics.

The coexistence of structural chirality and Néel-vector-driven polarization imposes a specific hierarchy of constraints on the symmetry operations of the system. Specifically, structural chirality requires the absence of space inversion ($\mathcal{P}$), mirror ($\mathcal{M}$), or roto-inversion ($\mathcal{\bar{S}}$) symmetries. Meanwhile, space inversion symmetry breaking is also a prerequisite for generating an electric polarization that can be tuned by reorienting the Néel vector. Although spin-orbit coupling (SOC) implies that Néel vector rotation leads to a redistribution of charge, the rotation itself does not inherently break $\mathcal{P}$ symmetry. Therefore, if the system preserves space inversion, the net polarization remains identically zero regardless of the magnetic orientation. By contrast, other rotation or mirror operations may be broken as the Néel vector aligns along different directions, thereby inducing a variation in polarization. Crucially, while all AMs are characterized by the breaking of the  [$\mathcal{C}_{2}||\mathcal{P}$] operation—which connects two sublattices with opposite spins—they may still retain the $[E||\mathcal{P}]$ operation that relates sublattices with the same spin orientation. Therefore, the essential symmetry criterion for realizing chiral altermagnetic magnetoelectrics with Néel-vector-driven polarization is the simultaneous absence of both [$\mathcal{C}_{2}||\mathcal{P}$] and $[E||\mathcal{P}]$ symmetry operations.

To further elucidate the microscopic physical mechanisms underlying chiral altermagnetic magnetoelectrics, we construct a representative tight-binding (TB) model that captures the electronic structures and the magnetoelectric coupling properties of the system [see the details  in the Supporting Information (\textcolor{blue}{SI})]. This model incorporates $p_z$ orbitals at the $2c$ and $6g$ Wyckoff positions of a hexagonal lattice with the magnetic space group $P6_3'22'$. The sublattices at the $2c$ positions host antiparallel magnetic moments, while the $6g$ sites remain nonmagnetic. Specifically, the opposite-spin sublattices are related by the $\{\mathcal{C}_{6z} | (0, 0, 1/2)\}$ screw operation, whereas the same-spin sublattices are connected by $\mathcal{C}_{3z}$ and $\mathcal{C}_{2x}$ rotations, naturally giving rise to the coexistence of $g$-wave altermagnetism and structural chirality. To investigate the chirality dependence, we further construct an enantiomorphic partner related by a mirror operation $\mathcal{M}_{x}$, denoted as the $C_L$ and $C_R$ models, respectively. The calculated band structures in Figures \ref{fig1}(a) and \ref{fig1}(b) exhibit characteristic $g$-wave altermagnetic features, where spin splitting emerges along the H-K-H$^\prime$ high-symmetry path and exhibits opposite signs in the two enantiomers.

Upon incorporating SOC, the electric polarization $P$ becomes explicitly dependent on the orientation of the Néel vector. Taking the $C_L$ model as an example: when the Néel vector is aligned along the $z$-axis, the in-plane polarization vanishes due to the preserved in-plane $\mathcal{C}_3$ symmetry, despite the absence of space inversion. Wilson loop calculations at $k_z=0$ confirm that the charge centers are pinned at the lattice centers under this configuration. However, reorienting the Néel vector away from the $z$-axis breaks the $\mathcal{C}_3$ symmetry, thereby lifting the symmetry constraints on polarization. Specifically, rotating the Néel vector toward the $x$-direction induces a shift of the charge center, resulting in a finite in-plane polarization. Conversely, a rotation toward the $-x$-direction reverses the charge displacement and the corresponding polarization [Figure \ref{fig1}(c)]. For the $C_R$ model, while the polarization is likewise zero for a $z$-aligned Néel vector, its rotation induces a polarization with the same magnitude and opposite sign to that in the $C_L$ model [Figure \ref{fig1}(d)].

\begin{figure}[t]
\centering
\includegraphics[width=1\linewidth]{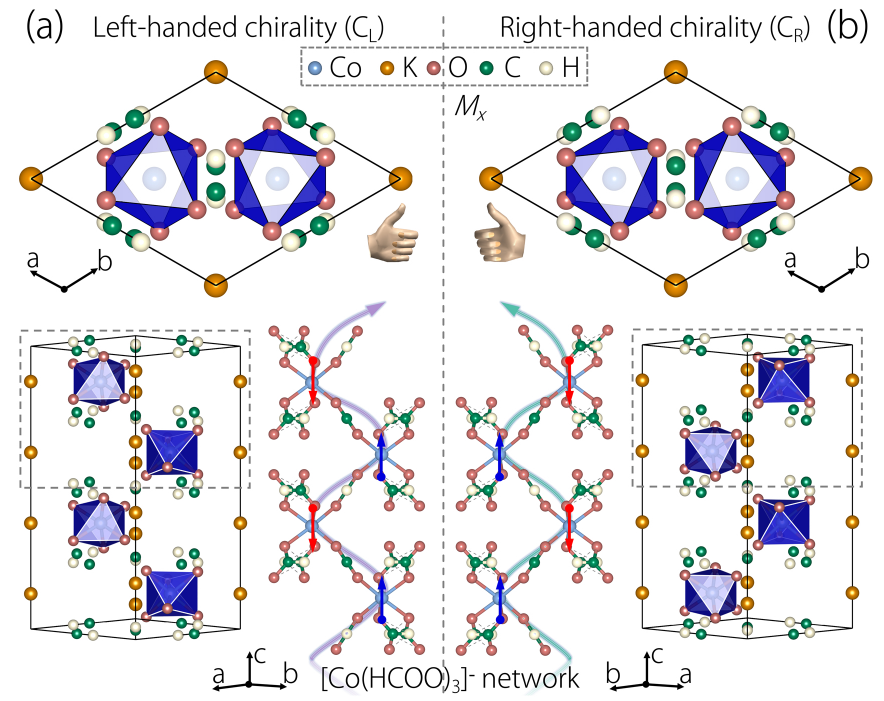}
\caption{Crystal structures of the (a) left-handed (C$_L$) and (b) right-handed (C$_R$) enantiomers for the 3D chiral AM K[Co(HCOO)$_3$], showing the top view, side view, and chiral framework of the [Co(HCOO)$_3$]$^{-}$ network. Among them, the red and blue arrows denote opposite spins, and the gray dashed squares indicate the unit cell.
\label{fig2}}
\end{figure}

\textit{\textcolor{blue}{Emergent chiral altermagnetism in K[Co(HCOO)$_3$]---}}Chiral altermagnetism is a special subclass of altermagnetism, wherein the alternating spin splitting can be switched by reversing the structural chirality of the lattice~\cite{add13}---analogous to multiferroic altermagnetism, where the spin splitting can be controlled via ferroelectric polarization~\cite{add19,add26,YQ-Zhu2025,MQ-Gu2025,ZY-Zhu2025}. Although chiral AMs are of considerable fundamental interest, their experimental realization remains elusive, mainly because suitable candidate materials are extremely scarce. The quasi-2D chiral compounds $X$Nb(Ta)$_3$S$_6$ ($X$ = V, Cr, Mn, Fe, Co, Ni)~\cite{Anzenhofer1970,Parkin1980} have emerged as promising chiral candidates; however, they are typically either conventional collinear AFMs~\cite{Parkin1983}, noncoplanar AFMs~\cite{Takagi2023}, or FMs~\cite{Parkin1980}, leaving only a handful of plausible cases (such as the V-based member)~\cite{add1,HY-Zhu2025}. Alternatively, we show that the MOF K[Co(HCOO)$_3$] provides a rare platform for 3D chiral altermagnetism. This compound has been demonstrated to exhibit antiferromagnetic order below a N{\'e}el temperature $T_N = 8.3$ K~\cite{add50}. Remarkably, single crystals of both C$_L$ and C$_R$ enantiomers can be grown separately, with the corresponding crystal structure data deposited under Cambridge Crystallographic Data Centre (CCDC) numbers 784051 and 784052~\cite{addCCDC}. These properties make the 3D MOF K[Co(HCOO)$_3$] a promising platform for the experimental realization of chiral altermagnetism.

Figures~\ref{fig2}(a) and~\ref{fig2}(b) illustrate the chiral crystal structures of the C$_L$ and C$_R$ enantiomers for the 3D MOF K[Co(HCOO)$_3$], which belongs to the $\emph{P}$6$_3$22 chiral space group (No. 182) [see Table \textcolor{blue}{S1} in the Supporting Information (\textcolor{blue}{SI})]. The structure features helical chains of Co$^{2+}$ cations interconnected by bridging [(HCOO)$_3$]$^{3-}$ ligands, forming a 3D chiral framework. This arrangement generates 1D channels occupied by K$^+$ cations within the interstitial sites. Our calculations reveal that the 3D MOF K[Co(HCOO)$_3$] exhibits antiferromagnetic ordering [see Figure \textcolor{blue}{S2} in the \textcolor{blue}{SI}]. The magnetism originates exclusively from the Co$^{2+}$ cations, with each Co$^{2+}$ center carrying a magnetic moment of 2.61 $\mu_B$, which is in excellent agreement with the experimental results~\cite{add50}. Furthermore, in 3D MOF K[Co(HCOO)$_3$], the hexagonal spin sublattice formed by Co$^{2+}$ cations is linked via the $[\mathcal{C}_2||\mathcal{C}_{6z}t]$ symmetry operation rather than $[\mathcal{C}_2||\bar{E}]$ or $[\mathcal{C}_2||t]$ in conventional AFMs ($\mathcal{C}_2$ is twofold rotation in spin space, while $\mathcal{C}_{6z}$, $\bar{E}$, and $t$ are real-space rotation, inversion, and translation, respectively), demonstrating that the 3D MOF K[Co(HCOO)$_3$] meets the symmetry requirements for altermagnetism.

\begin{figure}[t]
\centering
\includegraphics[width=1\linewidth]{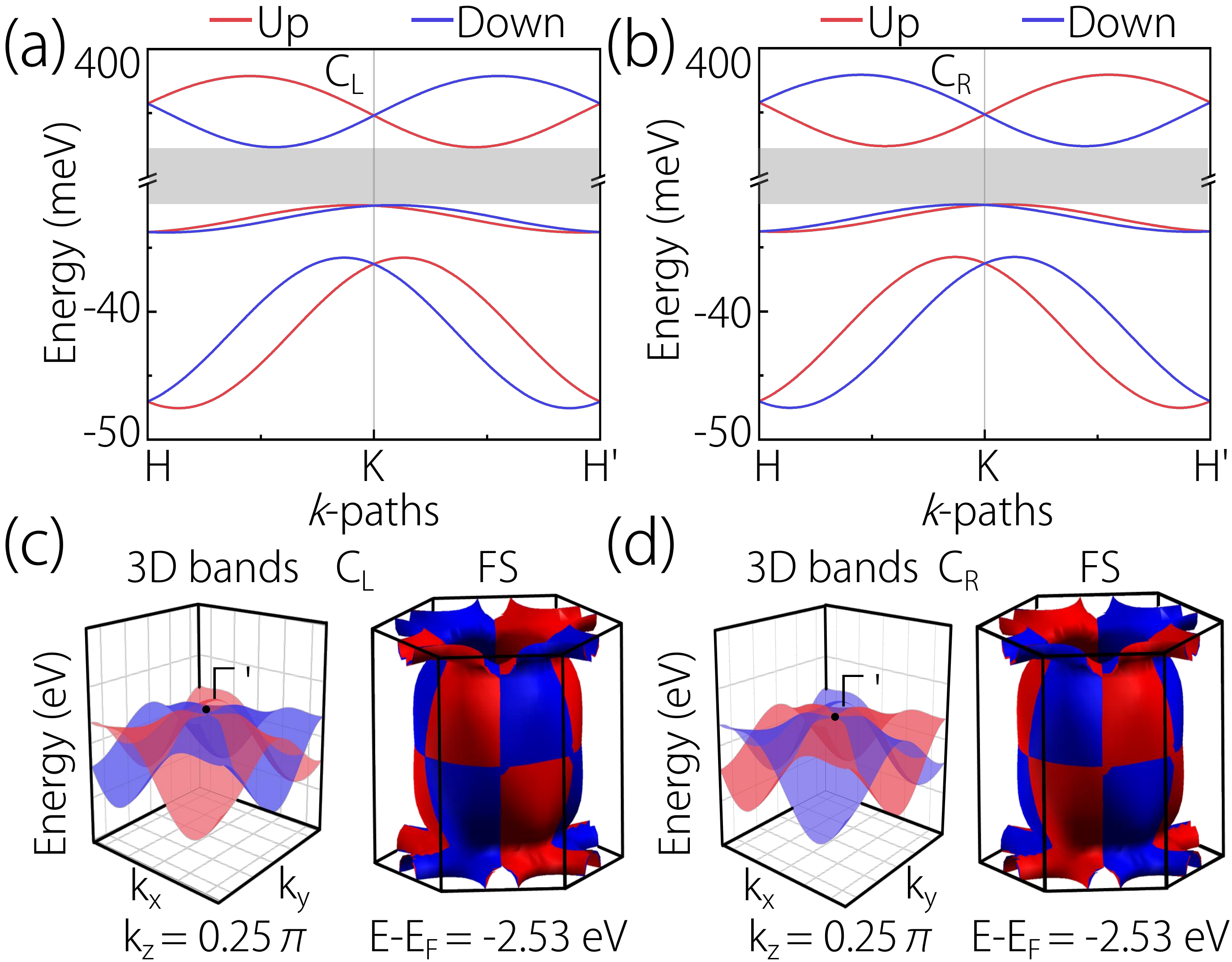}
\caption{(a) and (b) Electronic band structures along the high-symmetry paths H-K-H$^\prime$ for the C$_{L}$ and C$_{R}$ enantiomers of K[Co(HCOO)$_3$], respectively. (c) and (d) The 3D dispersion bands ($k_z$ = 0.25$\ \pi$) and Fermi surfaces (FSs) (E - E$_F$ = -2.53 eV) for the C$_L$ and C$_R$ enantiomers of K[Co(HCOO)$_3$], respectively.
\label{fig3}}
\end{figure}

Subsequently, the electronic band structures for the C$_L$ and C$_{R}$ enantiomers of the 3D MOF K[Co(HCOO)$_3$] are shown in Figure \textcolor{blue}{S3} in the \textcolor{blue}{SI}. Although band degeneracy is maintained along the high-symmetry paths $\Gamma$-M-K-$\Gamma$-A-L-H-A$|$L-M, pronounced spin splitting emerges along the H-K-H$^\prime$ direction, as shown in Figures~\ref{fig3}(a) and~\ref{fig3}(b). Notably, chiral switching induces complete spin-channel conversion (from $\uparrow$ to $\downarrow$) between the C$_L$ and C$_R$ enantiomers, a characteristic feature of chiral altermagnetism. Furthermore, the 3D band dispersion and Fermi surface reveal pronounced altermagnetic spin splitting, providing a direct visualization of the $g$-wave symmetry in momentum space, as shown in Figures~\ref{fig3}(c) and~\ref{fig3}(d).

\begin{figure}
\centering
\includegraphics[width=1\linewidth]{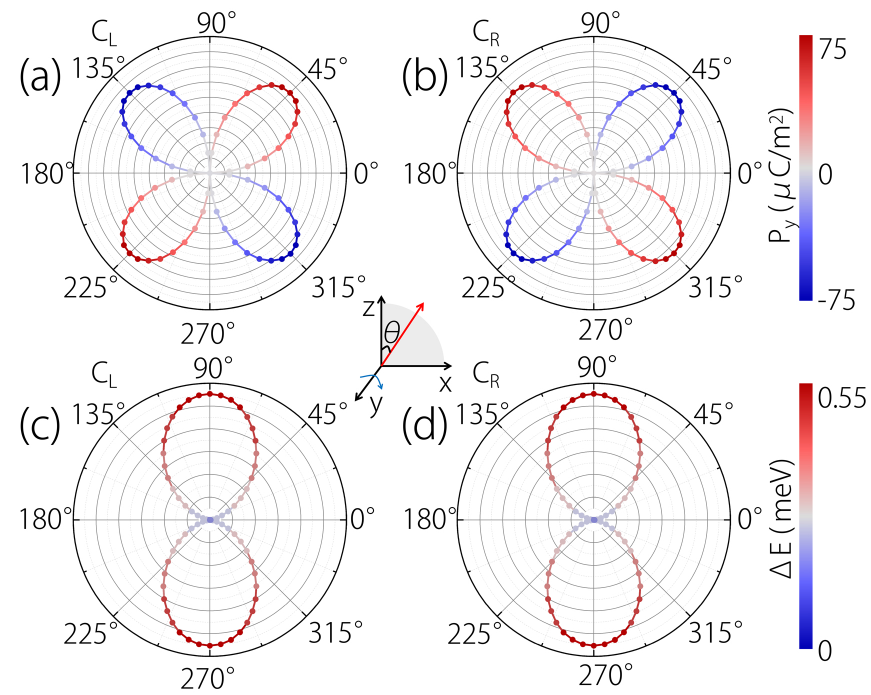}
\caption{$P_y$ as a function of the out-of-plane N\'eel-vector orientation $\theta$ within the $zx$ plane ($\theta=0^\circ$ corresponds to the crystallographic [001] direction). (a,b) Results for the C$_L$ and C$_R$ enantiomers, respectively. (c,d) Energy difference $\Delta E$ as a function of $\theta$ for the C$_L$ and C$_R$ enantiomers, respectively, with $\theta=0^\circ$ corresponds to the ground state and is taken as the reference.
\label{fig4}}
\end{figure}

\textit{\textcolor{blue}{Emergent altermagnetic magnetoelectricity in K[Co(HCOO)$_3$]---}}Next, we come to investigate the N\'eel-vector-driven polarization in the C$_L$ and C$_R$ enantiomers. With the Néel vector aligned along the $z(x)$ axis [$\theta = n\pi(\theta =\pi/2+n\pi)$] with $n\in\mathbb{N}$, the system belongs to the nonpolar magnetic point group $6'22'$ ($2'2'2$), which forbids any spontaneous electric polarization, and therefore represents a nonpolar AM (see Figure~\ref{fig4}). As the N\'eel vector rotates within the $zx$ plane away from the high-symmetry $z(x)$ direction, the magnetic symmetry is reduced from nonpolar magnetic point group to the polar magnetic point group $2'$ for a generic orientation. Under this symmetry, only the $P_y$ component is allowed and is  protected by the $\mathcal{TC}_{2y}$ symmetry, giving rise to a finite polarization perpendicular to the rotation plane. As shown in Figures~\ref{fig4}(a) and \ref{fig4}(b), first-principles calculations reveal a clear N\'eel-vector-driven polarization in both enantiomers. The polarization follows $P_y \propto \sin(2\theta)$, exhibiting a $180^\circ$ periodicity, while the two enantiomers carry exactly opposite signs. This behavior is strictly constrained by magnetic symmetry. For each fixed enantiomer, the states at $\theta$ and $180^\circ-\theta$ are related by $\mathcal{TC}_{2x}$ symmetry, under which $P_y$ is odd; therefore, they have equal magnitudes but opposite signs. By contrast, the states at $\theta$ and $180^\circ+\theta$ are related by time-reversal symmetry $\mathcal{T}$, under which $P_y$ is even, and therefore carry the same polarization. In addition, opposite chiral enantiomers are related by $\mathcal{TP}$ symmetry, under which $P_y$ is odd, so they exhibit equal-magnitude but opposite-sign polarization for the same N\'eel-vector orientation.

Notably, the maximum $|P_y|$ reaches $74.73~\mu\mathrm{C/m}^2$ at $\theta=\pi/4+n\pi/2$, exceeding those reported for collinear antiferromagnetic magnetoelectrics such as CuFeS$_2$ (${\sim}20~\mu\mathrm{C/m}^2$)~\cite{YX-Zhang2025} and V$X_2$ ($X=\mathrm{Cl, Br, I}$), whose values range from 12.2 to 64.8~$\mu\mathrm{C/m}^2$~\cite{C-Liu2024}. Moreover, as the N\'eel vector rotates from $0^\circ$ to $45^\circ$, $P_y$ evolves from zero to its maximum value across a small energy barrier of only 0.259 meV [see Figures~\ref{fig4}(c) and ~\ref{fig4}(d)], indicating that the polarization can be efficiently controlled through N\'eel-vector reorientation. Experimentally, the orientation of the N\'eel vector in AMs can be tuned by magnetic fields~\cite{C-Song2025}.
Recent experiments have demonstrated that 180$^\circ$ switching of the N\'eel vector can be achieved in AMs. Specifically, current-induced spin-orbit torque, assisted by a small magnetic field, has been shown to induce 180$^\circ$ switching of the N\'eel vector in Mn$_5$Si$_3$~\cite{L-Han2024b}, with the anomalous Hall effect serving as a detection tool. 

\begin{figure}[t]
\centering
\includegraphics[width=1\linewidth]{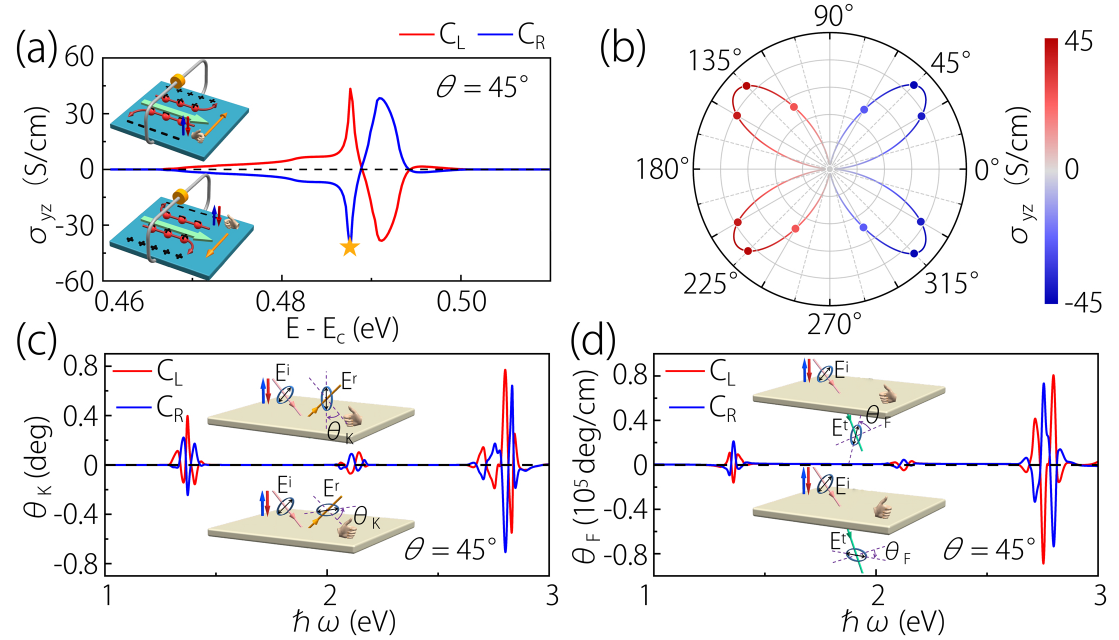}
\caption{Dual sign-reversible anomalous transport properties in K[Co(HCOO)$_3$]. (a) Anomalous Hall conductivity $\sigma_{yz}$ for the C$_L$ and C$_R$ enantiomers. The Fermi levels $E-E_c$ are aligned with the conduction-band minimum. The star marks the representative peak energy at which the angular dependence in (b) is evaluated. (b) Angular dependence of $\sigma_{yz}$ as a function of the N\'eel-vector orientation $\theta$ within the $zx$ plane. (c) Magneto-optical Kerr rotation angle $\theta_K$ and (d) Faraday rotation angle $\theta_F$ for the C$_L$ and C$_R$ enantiomers. In (a), (c), and (d), the N\'eel vector is oriented at $\theta=45^\circ$ within the $zx$ plane.
\label{fig5}}
\end{figure}

\textit{\textcolor{blue}{Electronic and optical probes of chiral magnetoelectric states---}}The same symmetry evolution that drives the magnetoelectric response also activates distinct electronic and optical signatures in K[Co(HCOO)$_3$], allowing the anomalous Hall and magneto-optical responses to serve as sensitive probes of the underlying chiral and N{\'e}el-vector states. When the N\'eel vector is aligned along the $z$ axis, the magnetic point group is $6'22'$, which forbids all anomalous Hall conductivity components. For a generic N{\'e}el-vector orientation within the $zx$ plane, however, the symmetry is reduced to the magnetic point group $2'$, under which both $\sigma_{yz}$ and $\sigma_{xy}$ are symmetry allowed. Therefore, the onset and sign evolution of these responses directly reflect the symmetry breaking induced by Néel-vector reorientation, providing effective readout channels for the switchable chiral magnetoelectric states.

To illustrate how these responses encode the underlying chiral and magnetic states, we consider the representative configuration at $\theta = 45^\circ$, for which the Hall response is symmetry allowed. Importantly, the two opposite chiral enantiomers are related by $\mathcal{TP}$ symmetry, under which the Hall response is odd. Therefore, opposite chiralities carry opposite $\sigma_{yz}$ for the same N{\'e}el-vector orientation, as shown in Figure~\ref{fig5}(a). In this sense, the anomalous Hall response provides a direct electronic fingerprint of lattice chirality in the chiral altermagnetic regime.

Besides distinguishing opposite chiralities, the anomalous Hall response also serves as a sensitive probe of N{\'e}el-vector rotation within a fixed enantiomer. Its angular evolution is strictly dictated by symmetry (see Table \textcolor{blue}{S3} in the \textcolor{blue}{SI}). The states at $\theta$ and $180^\circ-\theta$ are related by $\mathcal{T}C_{2x}$, under which both the polarization and the Hall response are odd. They therefore form a pair of symmetry-related states with opposite polarization and opposite transport signals ($\sigma_{yz}$), but equal magnitudes, offering potential advantages for device applications such as nonvolatile binary-state encoding and robust electrical readout~\cite{Kosub2017}. Moreover, $\sigma_{yz}$ also reverses sign for the states at $\theta$ and $180^\circ+\theta$ due to the odd property of Hall effect under $\mathcal{T}$ symmetry. These symmetry relations fully determine the sign evolution of the transport responses during N\'eel-vector rotation in the $zx$ plane, as summarized in Figure~\ref{fig5}(b).

The same symmetry-governed sign reversal further extends from the dc anomalous Hall response to magneto-optical probes in the ac regime (see Figure~\textcolor{blue}{S4} in the~\textcolor{blue}{SI}), so that the Kerr and Faraday rotation angles also reverse sign between opposite enantiomers. As shown in Figures~\ref{fig5}(c) and \ref{fig5}(d), both rotation angles exhibit mirrored spectral features for opposite chiralities, providing direct optical fingerprints of chirality-locked altermagnetic states. Quantitatively, the maximum Kerr and Faraday rotation angles in the 3D MOF K[Co(HCOO)$_3$] reach 0.77 deg and 0.89$\times10^5$ deg/cm, respectively [see Figures~\ref{fig5}(c) and~\ref{fig5}(d)]. Notably, the Kerr rotation angle is comparable or even exceeds those of several representative traditional FMs, non-collinear AFMs, and AMs, including fcc Ni ($\sim$0.15 deg)~\cite{Engen1983}, Mn$_3Y$ ($Y$ = Ge, Ga, Sn) ($<$ 0.02 deg)~\cite{Higo2018}, Mn$_3X$N ($X$ = Ga, Ni, Ag, Ni) ($<$ 0.42 deg)~\cite{XD-Zhou2019a}, hcp Co ($\sim$0.48 deg)~\cite{GY-Guo1994}, and RuO$_2$ ($\sim$0.62 deg)~\cite{add58}. These mirrored Kerr and Faraday spectra echo the sign-reversal behavior of the anomalous Hall response, confirming their common chiral origin. Together, the anomalous Hall and magneto-optical responses provide complementary electronic and optical probes of the coupled chirality and Néel-vector states in K[Co(HCOO)$_3$].

Finally, we briefly discuss the experimental feasibility of preparing, switching, and probing the chiral magnetoelectric states in the 3D MOF K[Co(HCOO)$_3$] based on existing technologies. Selective chiral-domain growth (C$_L$ or C$_R$ enantiomer) can be achieved using the self-flux method with a chiral seed crystal, as demonstrated in the chiral topological semimetal PdGa~\cite{add63} or driven in situ by circularly polarized light irradiation~\cite{add64,add65}. Once synthesized, chiral switching between the two enantiomers may be triggered by several established methods, such as photoexcitation-induced ligand rearrangements~\cite{add66,add67,add68,add69,add70}, electron tunneling~\cite{add71}, electric-field control~\cite{add72} or temperature-driven phase inversion~\cite{add73}. The resulting lattice chirality can be sensitively characterized by the x-ray diffraction with the Flack method~\cite{add63}, circular photogalvanic effect~\cite{add74,add75}, or scanning transmission electron microscopy~\cite{add76}. Our results further suggest that the anomalous Hall effect  and magneto-optical effects provide two complementary probe channels for resolving both lattice chirality and Néel vector orientation in K[Co(HCOO)$_3$], through their characteristic sign-locking behavior for different enantiomers and magnetic states. These chirality- and Néel-vector-dependent signatures can therefore serve as effective electronic and optical readouts of the underlying chiral altermagnetic magnetoelectric states, complementing established probes of altermagnetic order such as spin- and angle-resolved photoemission spectroscopy~\cite{add7,add77}, as recently demonstrated by  anomalous Hall effect  and magneto-optical effects measurements in AMs RuO$_2$, MnTe, CrSb, and FeS~\cite{add14,add78,add80,add81,add82,add83,add84}. Remarkably, high-quality single crystals of MOF K[Co(HCOO)$_3$], encompassing both C$_L$ and C$_R$ enantiomers, have already been successfully synthesized~\cite{add50}, providing an experimentally accessible platform for detecting these chirality- and Néel-vector-resolved probe signals in the altermagnetic regime. The feasibility of selective chiral-state preparation, reversible chirality tuning, and multi-channel electrical and optical detection highlights the experimental accessibility of chiral altermagnetic magnetoelectricity in K[Co(HCOO)$_3$] and points to a promising route toward chirality-programmable nonvolatile spintronic functionalities.

\textit{\textcolor{blue}{Conclusion---}}We theoretically uncover a previously unexplored form of chiral altermagnetic magnetoelectrics and, based on a general theoretical model, reveal a universal mechanism by which structural chirality and N\'eel-vector-driven polarization become intrinsically coupled in nonpolar altermagnetic systems. Using the 3D MOF K[Co(HCOO)$_3$] as a representative material platform, we further show that reorientation of the N\'eel vector induces a finite electric polarization in this otherwise nonpolar AM, while switching between opposite chiral enantiomers produces an additional sign reversal, thereby enabling dual-mode control of the magnetoelectric state by structural chirality and N\'eel-vector orientation. The coupled chiral and magnetic states further give rise to distinct electronic and optical probe signatures, including sign-reversible anomalous Hall and magneto-optical responses, which provide effective readout channels for both chirality and N\'eel-vector configurations. Crucially, high-quality single crystals of both left- and right-handed enantiomers of 3D MOF K[Co(HCOO)$_3$] have already been synthesized, and its antiferromagnetic order has been experimentally established, offering an experimentally accessible and tunable platform. Our work establishes K[Co(HCOO)$_3$] as a prototype chiral altermagnetic magnetoelectric and opens a promising route toward chirality- and N\'eel-vector-controlled magnetoelectric functionalities and probe schemes in altermagnetic materials.

\textit{\textcolor{blue}{Acknowledgments---}}This work is supported by the National Key R$\&$D Program of China (Grant No. 2022YFA1402600), the Australian Research Council Discovery Early Career Researcher Award (Grant No. DE240100627), the Australian Research Council Discovery Project (Grant No. DP260102992), the National Natural Science Foundation of China (Grant No. 12304066, No. 12404073), the Hebei Natural Science Foundation (Grant No. A2024205006), and the Basic Research Program of Jiangsu (Grant No. BK20230684). We also acknowledge the computational resources from the National Computational Infrastructure (NCI), which were allocated from the National Computational Merit Allocation Scheme supported by the Australian Government.

\bibliography{ref}

\end{document}